\documentclass[letterpaper,prl,10pt,twocolumn,showpacs,showkeys,preprintnumbers,superscriptaddress,amsmath,amssymb,nofootinbib,notitlepage]{revtex4-1}
\usepackage{graphicx, epsfig, amssymb} 
\usepackage{amsmath, amsfonts}
\usepackage{bm} 
\usepackage{color}
\usepackage[usenames]{xcolor}
\usepackage{hyperref}
\usepackage{siunitx}
\hypersetup{linktocpage,colorlinks=true,urlcolor=blue,linkcolor=blue,citecolor=blue}  
\usepackage{verbatim}
\usepackage[utf8]{inputenc}
\usepackage{natbib}
\usepackage{bbm}

	\newcommand{\ncd}{\newcommand}
	\ncd{\mrm}    {\mathrm}
	\ncd{\beq} {\begin{equation}}
	\ncd{\eeq} {\end{equation}}
	\ncd{\nn}{\nonumber}

	\def\d{{\rm d}}

	\def\basis[#1]{\frac{\partial}{\partial #1}}
	\def\dt[#1]{\frac{\d}{\d #1}}

	\def\ric{{\rm Ric}}

	\newtheorem{corola-ry}{Corolary}

\begin{document}
 
\title{Thurston Geometries in Three-Dimensional New Massive Gravity}
\author{Daniel \surname{Flores-Alfonso}}
\email[]{daniel.flores@cinvestav.mx}
\affiliation{Departamento de F\'isica, CINVESTAV-IPN, A.P. 14-740, C.P. 07000, Ciudad de México, Mexico}

\author{Cesar S.~\surname{Lopez-Monsalvo}}
\email[]{cslopezmo@conacyt.mx}
\affiliation{Conacyt-Universidad Aut\'onoma Metropolitana Azcapotzalco,  Avenida San Pablo Xalpa 180, Azcapotzalco, Reynosa Tamaulipas, C.P. 02200, Ciudad de M\'exico, Mexico}

\author{Marco \surname{Maceda}}
\email[]{mmac@xanum.uam.mx}
\affiliation{Departamento de F\'isica,
Universidad Aut\'onoma Metropolitana - Iztapalapa,
Avenida San Rafael Atlixco 186, A.P. 55534, C.P. 09340, Ciudad de M\'exico, Mexico}


\begin{abstract}
We show that the three-dimensional Thurston geometries are vacuum solutions to the 3D new massive
gravity equations of motion. We analyze their Lorentzian counterparts as well.
\end{abstract}

\maketitle

\emph{Introduction.}---Thurston’s conjecture is a statement about how any three-dimensional manifold can be canonically decomposed into parts. These simpler parts admit one, and only one, of eight model geometries. Shortly after Thurston’s conjecture~\cite{Thurston:1982}, Ricci flow was created as an attempt to prove it~\cite{Hamilton:1982}. The Ricci flow equation has the same form as a heat equation, in this analogy, the role of the temperature is played by the metric. The first sketch of proof that the conjecture is true came decades later~\cite{Perelman}.

In gravitational physics, Thurston geometries enter the picture in 4D spatially homogeneous cosmological models~\cite{Taub:1950ez,Kantowski:1966te,Fagundes:1985} and have been used to construct 5D black holes~\cite{Cadeau:2000tj,Hassaine:2015ifa,Bravo-Gaete:2017nkp}. In 3D gravity, Thurston metrics are interpreted as gravitational instantons. Some Thurston model geometries are Euclidean vacua of massive gravity theories (see Ref.~\cite{Chernicoff:2018hpb} for a recent discussion on some aspects of exotic massive gravity~\cite{Ozkan:2018cxj}).
Let us mention that there are a number of massive gravity theories in three dimensions, many of which are closely related to
topologically massive gravity (TMG)~\cite{Deser:1981wh,Deser:1982vy}; some of these relations are discussed in~\cite{Ozkan:2018cxj}.
However, no systematic study has been undertaken so far.
In this Letter, we address the question if all Thurston geometries are vacuum solutions of three-dimensional new massive gravity (NMG)~\cite{Bergshoeff:2009hq}.

We are concerned with the following equations of motion
\begin{equation}
\Lambda g+ G(g)-\frac{1}{2m^2}K(g)=0. 
\label{EOM}
\end{equation}
Here $\Lambda$ is the cosmological constant and we parameterize the NMG coupling constant as $m^{-2}$, however, there is no need for $m^2$ to be positive.
The NMG tensor, $K$, has components
\begin{align}
K_{\mu\nu}=&~2\square {R}_{\mu\nu}-\frac{1}{2}\nabla_{\mu}
\nabla_{\nu }{R}-\frac{1}{2}\square {R}g_{\mu\nu}
+4R_{\mu\alpha\nu\beta}R^{\alpha\beta}\notag\\
&-\frac{3}{2}RR_{\mu\nu}-R_{\alpha\beta}R^{\alpha\beta}g_{\mu\nu}
+\frac{3}{8}R^{2}g_{\mu\nu}.
\label{KNMG}
\end{align}

It was established, in a sequence of papers~\cite{Ahmedov:2010em,Ahmedov:2010uk,Ahmedov:2011yd}, that Eq. \eqref{EOM} admits the form of a Klein-Gordon-type equation with a curvature-squared source term and a constraint equation. This ultimately allows for all algebraic type $D$ and $N$ solutions of TMG to be mapped into NMG. The reason behind this map is that for these algebraically special metrics the TMG equations of motion correspond to a ``square root'' of the NMG Klein-Gordon-type equation. In contrast,
the converse is not true, that is, not every type $D$ and $N$ solution of NMG is also a solution of TMG.

In addition, the study of the various compatible geometric structures on three-dimensional manifolds has been
a central theme in contact geometry.
Recently, a class of contact structures has  emerged, arising in the context of six-dimensional supergravity~\cite{Murcia:2019cck}, three-dimensional massive gravity and two-dimensional superconductors~\cite{Flores-Alfonso:2020zif}. They have been dubbed $\varepsilon$-contact manifolds, where the sign of $\varepsilon$ indicates if a distinguished vector field in the manifold is timelike, spacelike or null. 
For example, the contact sphere with a para-Sasakian structure is one such manifold where 
the Maxwell-Chern-Simons field equations are obtained by extremizing its helicity while identifying the contact form with the electromagnetic gauge potential.

Previous results show that $\varepsilon$-contact three-manifolds that are also $K$-contact and $\eta$-Einstein are solutions of NMG. In Ref.~\cite{Flores-Alfonso:2020zif} we specialized in a Lorentzian Nil geometry~\cite{Rahmani}. However, our results are independent of the metric's signature. In particular, every Sasakian geometry satisfies the NMG equations of motion. This should be understood in light of the uniformization theorem of Sasakian three-manifolds~\cite{Belgun:2000,Boyer:2004eh}. The three representative metrics of the theorem are all Thurston model geometries. 

\emph{Thurston Geometries.}---In this section, we summarize and discuss 
the results of evaluating the NMG equations of motion in each of the Thurston geometries: $E^3$, $S^3$, $H^3$, $E^1\times S^2$, $E^1\times H^2$, $\widetilde{{\rm{SL}}(2,\mathbb{R})}$, Nil, and Sol~\cite{Thurston:1980,Thurston,Scott:1983}.

Let us consider first, the three Thurston geometries formed by a direct product $E^1\times E^2=E^3$, $E^1\times S^2$ and $E^1\times H^2$. Euclidean space $E^3$ is flat and thus to be a solution of NMG it only requires
\begin{equation}
 \Lambda=0,
\end{equation}
in other words $m^2$ is unspecified. For the remaining product spaces we write the scalar curvature as
\begin{equation}
 R=\pm\frac{2}{\ell^2},
\end{equation}
where the plus sign is for $E^1\times S^2$. As NMG vacua the length scale $\ell$ is fixed by the cosmological constant. Moreover, the quadratic-curvature coupling constant is also fixed by $\Lambda$
\begin{equation}
 \ell^2=\pm\frac{1}{2\Lambda},\quad m^2=\Lambda,
 \label{E1xH2}
\end{equation}
where, once more, the plus sign is for $E^1\times S^2$. Lorentzian counterparts of these spaces are (A)dS$_2\times S^1$ which 
were found to be a solution of Eq. \eqref{EOM} in Refs.~\cite{Clement:2009gq,Ahmedov:2011yd}, as long as Eqs. \eqref{E1xH2} hold.

Although Euclidean space $E^3$ is a product space $E^1\times E^2$ it is also a space of constant (sectional) curvature $\kappa$.
Evidently, $\kappa$ vanishes for $E^3$.
The other two spaces with constant $\kappa$ are $S^3$ ($\kappa>0$) and $H^3$ ($\kappa<0$).
We write their scalar curvature as
\begin{equation}
 R=\pm\frac{6}{\ell^2},
\end{equation}
where the minus sign is for $H^3$. As it is for $E^3$, $m^2$ is unfixed by the NMG equations of motion, however there is a constraint in the parameter space given by
\begin{equation}
 \Lambda=\pm\frac{1}{\ell^2}-\frac{1}{4\ell^4m^2}. \label{S3H3}
\end{equation}
This equation determines the value of $\ell$ in terms of the NMG couplings and corresponds to Eq. (3.9) of Ref.~\cite{Bakas:2010kc} for the plus sign.
Moreover, these are the same equations (A)dS spacetimes must satisfy as NMG vacua, cf. \cite{Bergshoeff:2009hq,Liu:2009kc,AyonBeato:2009yq}. However, since $\ell^2$ must be positive then the range of values of the coupling constants are restricted. Notice that when $m^{-2}=0$ we recover Einstein gravity and the familiar conditions $\ell^2=\pm\Lambda^{-1}$.

Thus far, we have discussed five of the eight Thurston model geometries. These five geometries are closely related to the three representative geometries of the uniformization theorem for Riemann surfaces, i.e., $H^2$, $E^2$, and $S^2$. We considered the trio of product spaces and the trio with constant sectional curvature, $\kappa$, where $E^3$ appears in both triplets. We now advance by looking at yet another trio $\widetilde{{\rm{SL}}(2,\mathbb{R})}$, Nil, and $S^3$ which are all line bundles over the representative Riemann surfaces mentioned above.

This new triplet appears in a version of the Uniformization Theorem that exists in three dimensions for Sasakian manifolds~\cite{Belgun:2000}. In short, compact Sasakian three-manifolds admit only three different types of geometries. The representatives of these three types are precisely Nil, $S^3$, and $\widetilde{{\rm{SL}}(2,\mathbb{R})}$. Furthermore, these correspond to the only three Thurston models on compact Seifert bundles with nonzero Euler number~\cite{Scott:1983}. They are all contact manifolds satisfying the $\eta$-Einstein condition~\cite{okumura1962}
\begin{equation}
 \ric(g)=\lambda g+\nu\eta\otimes\eta,
 \label{etaEin}
\end{equation}
where $\eta$ is the contact form, $g$ is the metric, and $\lambda,\nu$ are some constants.
When $\nu=0$ the manifold is also an Einstein manifold, this is the case for $S^3$. Just as three-manifolds are Einstein manifolds if and only if they have constant $\kappa$, Sasakian three-manifolds are $\eta$-Einstein if and only if they have
constant $\phi$-holomorphic sectional curvature $k$~\cite{okumura1962,tanno1969,Boyer:2004eh}.

Now, let us recall that a contact manifold 
possesses a special hyperplane distribution in its tangent bundle. In three-manifolds this contact distribution is a plane distribution. We mention this now because sectional curvature is a function of plane distributions. In three-manifolds any tangent space is three dimensional. Thus, there are at each point of the manifold $\binom{3}{2}=3$ independent planes. We say that a manifold has constant sectional curvature if for any plane at any point $\kappa$ takes the same value. In Sasakian manifolds, we say that a manifold has constant $\phi$-holomorphic sectional curvature if the sectional curvature restricted to the contact distribution is constant~\cite{tanno1969}. Here $\phi$ refers to the almost contact structure. Every plane on the distribution is generated by some vector field $X$ on it and $\phi X$, which results from the action of $\phi$ on $X$. The point is that manifolds such as Nil and $\widetilde{{\rm{SL}}(2,\mathbb{R})}$ do not have constant sectional curvature but do have constant $\phi$-holomorphic sectional curvature. 

Let us mention that every Sasakian manifold is $\varepsilon$-contact and $K$-contact. Thus, by our previous results~\cite{Flores-Alfonso:2020zif}, the $\eta$-Einstein geometries described above are vacuum solutions of NMG. Let us discuss this in further detail starting with the Nil geometry which is described by the metric
\begin{equation}
 g=\frac{\ell^2}{4}\bigg(\d x\otimes\d x +\d y\otimes \d y+\eta\otimes\eta\bigg), \label{Nil}
\end{equation}
where
\begin{equation}
 \eta=x\d y-\d z. \label{etaNil}
\end{equation}
In this case, the sectional curvature of planes in the contact distribution is $k=-3\ell^{-2}$. In NMG, the length scale $\ell$ is set by the cosmological constant by
\begin{equation}
 \ell^2=-\frac{1}{2\Lambda},
\end{equation}
and the NMG parameters must obey
\begin{equation}
m^2=21\Lambda<0,
\end{equation}
in order for \eqref{Nil} to satisfy the equations of motion.

We now turn to the universal cover of ${\rm{SL}}(2,\mathbb{R})$, which is described by the metric
\begin{equation}
 g=\ell^2\big(\d r\otimes\d r+\sinh^2r\cosh^2r\d\theta\otimes\d\theta
 +\eta\otimes\eta\big), \label{SL2R}
\end{equation}
with
\begin{equation}
 \eta=\d\psi+\sinh^2r\d\theta. \label{etaSL2R}
\end{equation}
The Sasakian structure on $\widetilde{{\rm{SL}}(2,\mathbb{R})}$ includes the (1,1) tensor
\begin{align}
\phi=&-\sinh r\cosh r\frac{\partial}{\partial r}\otimes\d\theta\notag\\
&+\frac{1}{\sinh r\cosh r}\frac{\partial}{\partial \theta}\otimes\d r-\frac{\sinh r}{\cosh r}\frac{\partial}{\partial \psi}\otimes\d r.
\end{align}
Additionally, let us mention that the $\phi$-holomorphic sectional curvature of the metric is $k=-7\ell^{-2}$. Moreover, inserting Eq. \eqref{SL2R} into the NMG equations of motion, Eq. \eqref{EOM}, yields
\begin{equation}
 \ell^2=-\frac{109}{50\Lambda},\quad\text{and}\quad m^2=\frac{625}{109}\Lambda<0,
 \label{universalsl2r}
\end{equation}
which tell us how the cosmological constant fixes the curvature radius and how the coupling constants are constrained so that the geometry of $\widetilde{{\rm{SL}}(2,\mathbb{R})}$ is a vacuum solution.

In sum, except for Sol, all Thurston geometries are either a space with a product structure, constant sectional curvature or constant $\phi$-holomorphic sectional curvature. For the Sol geometry we write its metric as
\begin{equation}
 g=\ell^2\big(e^{-2z}\d x\otimes\d x +e^{2z}\d y\otimes\d y+\d z\otimes\d z\big).
 \label{sol}
\end{equation}
The isometry algebra of this metric is solvable.

Considering now the NMG equations, we find that the conformal scale is fixed by
\begin{equation}
 \ell^2=-\frac{1}{2\Lambda}, \label{sol1}
\end{equation}
telling us that $\Lambda$ must be negative and, what is more,
\begin{equation}
 m^2=5\Lambda. \label{sol2}
\end{equation} 

Anticipating the results provided below let us note that the Jordan normal form of the Cotton-York tensor for the Sol geometry is
\begin{equation}
Y =2l^3
\left(\begin{array}{ccc}
-1 & 0 & 0 \\
 0 & 0 & 0 \\
 0 & 0 & 1
\end{array}\right),
\label{Jsol}
\end{equation}
which tells us that it has two independent real eigenvalues. Thus, the Sol geometry is algebraically general in the classification of~\cite{Garcia:2003bw}.

\emph{Lorentzian Signature.}---Since NMG is a gravity theory, obviously, one is interested in Lorentzian geometries.
Some Lorentzian metrics are obtained from Euclidean ones by a Wick rotation~\cite{Wick:1954eu}. For instance Minkowski from $E^3$, de Sitter from $S^3$, and anti-de Sitter (AdS) from $H^3$ or $\widetilde{{\rm{SL}}(2,\mathbb{R})}$. However, since there is no distinguishable direction over which to perform the rotation other inequivalent metrics arise from the same Euclidean geometry. The canonical Lorentzian metric on the three-sphere is an example
\begin{equation}
 g=\frac{\ell^2}{4}\bigg(-\eta\otimes\eta+\d\Theta\otimes\d\Theta+\sin^2\Theta\d\Phi\otimes\d\Phi\bigg), \label{Lorentz-SU2}
\end{equation}
where
\begin{equation}
 \eta=\d\Psi+\cos\Theta\d\Phi.\label{etaSU2}
\end{equation}
This metric is not diffeomorphic to de Sitter spacetime but both are obtained from Wick rotations applied to the Thurston geometry $S^3$.

The metric in Eq. \eqref{Lorentz-SU2} does not have constant sectional curvature. However, if we restrict ourselves to planes in the contact distribution then the sectional curvature of \eqref{Lorentz-SU2} is constant. This is
\begin{equation}
 \kappa(p)=\frac{7}{\ell^2}, \label{kappa(p)}
\end{equation}
for $p$ in the contact distribution generated by $\eta$. The metric is $\eta$-Einstein and $K$-contact. Moreover Eq. \eqref{kappa(p)} is highly suggestive it relates to $\widetilde{{\rm{SL}}(2,\mathbb{R})}$. Indeed, requiring the metric in Eq. \eqref{Lorentz-SU2} to satisfy the NMG equations of motion leads to
\begin{equation}
 \ell^2=\frac{109}{50\Lambda},\quad\text{and}\quad m^2=\frac{625}{109}\Lambda>0,
 \label{lorentziansu2}
\end{equation}
which should be compared with Eqs. \eqref{universalsl2r}.
Perhaps these results are not so surprising when one keeps in mind that SU(1,1) is isomorphic to ${\rm{SL}}(2,\mathbb{R})$.

The metric in Eq. \eqref{Lorentz-SU2} is a special case of the only SU(2) left-invariant family of Lorentzian geometries~\cite{Boucetta}. Except for $S^2\times E^1$, all Thurston geometries are left-invariant metrics. The para-Sasakian Nil geometry mentioned above belongs to one of two nonflat families of left-invariant Lorentzian Nil geometries. The other one is  
\begin{equation}
 g=\frac{\ell^2}{4}\bigg(\d x\otimes\d x +\d y\otimes \d y-\eta\otimes\eta\bigg), \label{perfectHeis}
\end{equation}
where $\eta$ is given by Eq. \eqref{etaNil}. This metric has been studied in the physics literature before, see, e.g., Ref.~\cite{Mukhopadhyay:2013gja} where the metric is studied in TMG and the Cotton-York tensor is interpreted as the energy-momentum tensor of a perfect fluid.
This carries over to the case at hand as follows.
The $\varepsilon$-contact, $\eta$-Einstein, and $K$-contact properties of the metric and the contact form imply~\cite{Flores-Alfonso:2020zif}
\begin{equation}
 K=ag+b\eta\otimes\eta,
\end{equation}
for some constants $a$ and $b$.
A direct calculation shows Eq. \eqref{perfectHeis} is a solution of NMG. In fact, the restrictions on the parameter space are exactly as in Eq. \eqref{LHeis}. By defining a unit timelike form 
\begin{equation}
 u=\frac{l}{2}\eta,\label{ueta}
\end{equation}
we now have
\begin{equation}
 \frac{1}{2m^2}K=pg+(\rho+p)u\otimes u.
\end{equation}
In other words, on shell, the NMG equations are effectively those of Einstein gravity
coupled to a perfect fluid, where the isotropic pressure and energy density are given by 
\begin{equation}
 p=3\Lambda,\quad\text{and}\quad\rho=5\Lambda,
\end{equation}
respectively.

The geometry in Eq. \eqref{perfectHeis} and other left-invariant metrics on Nil, SU(2), and SU(1,1) have been studied before in the context of NMG, as Lorentzian Bianchi type metrics~\cite{Ahmedov:2010em,Ahmedov:2011yd}.
What is more, except for $H^3$ and $E^1\times H^2$, one or more Lorentzian analogs of the Thurston geometries have been studied in these works.

\emph{The Dumitrescu-Zeghib metrics.}---Thus far, we have seen
that some Lorentzian left-invariant metrics sharing isometry group with the Thurston metrics 
are NMG vacua. Some also lead to applications in physics.
Recalling that there is no Lorentzian counterpart to the correspondence between Riemannian metrics and reductions of the structure group of the frame bundle (cf. Theorem 5.8 in~\cite{isham1999modern}), the Lorentzian counterpart to the geometrization theorem remains an open problem.

Nevertheless, amongst the various possible Lorentzian analogs of the Thurston metrics, in this Letter we study those found in~\cite{Dumitrescu:2010}, where it was established that
there are exactly four left-invariant metrics which are Lorentzian and at the same time are preserved by the canonical action of the isometry group. This is compelling
because it is exhaustive and provides only a handful of metrics. More precisely, it was established that any compact locally
homogeneous Lorentz threefold $M$ is isometric to a quotient of a Lorentz homogeneous space
$G/I$ by a discrete subgroup $\Gamma$ of $G$ acting properly and freely on $G/I$. A famous, closely related, \emph{noncompact} spacetime is the Ba\~nados-Teitelboim-Zanelli (BTZ) black hole~\cite{Banados:1992wn}.
It is locally homogeneous and is isometric to a quotient of a Lorentz homogeneous space by a discrete subgroup~\cite{Banados:1992gq}. Specifically, it is AdS spacetime with additional (nonobvious) identifications; for further details we recommend~\cite{AyonBeato:2004if}. A further result of~\cite{Dumitrescu:2010} is that when $I$ is noncompact then $G/I$ is isometric to a Lie group $L$ endowed with a Lorentzian left invariant metric, where $L$ is isomorphic to $E^3$, $\widetilde{{\rm{SL}}(2,\mathbb{R})}$, Nil or Sol.

For $E^3$ and $\widetilde{{\rm{SL}}(2,\mathbb{R})}$ these are the Minkowski and AdS metrics, respectively. The Minkowski metric is one of only three left-invariant Lorentzian metrics on Nil~\cite{Rahmani}, however, it is not preserved by the canonical action of Nil. From Ref.~\cite{Boucetta} we know that there are in total seven left-invariant Lorentzian metrics on $\widetilde{{\rm{SL}}(2,\mathbb{R})}$. However, only the AdS metric is the one which is preserved by its canonical action.

The Minkowski metric is flat and thus the only requirement for it to be a NMG vacuum solution is $\Lambda=0$. Turning to AdS we see that it is analogous to two Thurston geometries. On the one hand, it has negative constant sectional curvature like $H^3$. This implies that for it to satisfy the NMG equations it must satisfy Eq. \eqref{S3H3} with the negative sign. On the other hand, AdS can also be obtained from Eq. \eqref{SL2R}, i.e.,
\begin{equation}
 g=\ell^2\big(\d r\otimes\d r+\sinh^2r\cosh^2r\d\theta\otimes\d\theta
 -\eta\otimes\eta\big),
\end{equation}
where $\eta$ is the same as in Eq. \eqref{etaSL2R}. This means that AdS spacetime is also analogous to the geometry of $\widetilde{{\rm{SL}}(2,\mathbb{R})}$.

The Nil group is also called the Heisenberg group because of its relation to quantum mechanics. In Ref.~\cite{Dumitrescu:2010} Lorentz-Heisenberg (Lorentz-Sol) is the name given to the unique left-invariant Lorentzian geometry which is preserved by the canonical action of the Nil (Sol) group. We use the same nomenclature in what follows.

We write the Lorentz-Heisenberg metric as
\begin{equation}
 g=\frac{\ell^2}{4}\bigg(-\d x\otimes\d x +\d y\otimes \d y+\eta\otimes\eta\bigg), \label{Lorentz-Heisenberg}
\end{equation}
where $\eta$ is as in Eq. \eqref{etaNil}. This geometry parallels its Euclidean counterpart. 
That is, it is para-Sasakian, $\eta$-Einstein and has constant $\varphi$-holomorphic sectional curvature $c=3\ell^{-2}$. Here $\varphi$ is the (1,1) tensor in the almost para-contact structure of the Nil group. It is a NMG vacuum under the constraints
\begin{equation}
 m^2=21\Lambda>0, \label{LHeis}
\end{equation}
and the cosmological constant is positive, as
\begin{equation}
 \ell^2=\frac{1}{2\Lambda}.
\end{equation}

Unlike the Lorentzian Nil geometries, there are Lorentzian left-invariant Sol metrics which cannot be obtained by (obvious) Wick rotations of Eq. \eqref{sol}. However, it has been proven that only seven such metrics exist~\cite{Boucetta}. The Lorentz-Sol metric is the only one among them which is conformally flat (``Petrov'' type $O$). We write the metric as
\begin{equation}
 g=\ell^2\big(2e^{-z}\d x\otimes_s\d z +e^{2z}\d y\otimes\d y\big),
 \label{Lorentz-Sol}
\end{equation}
where $A\otimes_sB=1/2\,(A\otimes B+B\otimes A)$. This metric is \emph{not} a solution of Eq. \eqref{EOM} since
\begin{equation}
 \ric(g)=-2\d z\otimes\d z,
\end{equation}
and the NMG tensor $K$ defined in Eq. \eqref{KNMG} vanishes. 
However, this means it is actually
a solution of pure new massive gravity,
where we have used the same nomenclature as in Pure Lovelock Gravity~\cite{Kastor:2006vw,Cai:2006pq}.
Hence, the configuration requires $\Lambda=0$ and leaves $m^2$ unspecified.

Although other left-invariant Sol metrics have been studied in NMG~\cite{Ahmedov:2010em}, the metric in Eq. \eqref{Lorentz-Sol} is a new solution. This motivated us to inspect the other Lorentzian Sol metrics. It was found that there are only three other metrics which satisfy the NMG equations of motion and we discuss them below.

\emph{New Sol geometries.}---We start with metric
\begin{equation}
 g_{I}=\ell^2\big(e^{-2z}\d x\otimes\d x +e^{2z}\d y\otimes\d y-\d z\otimes\d z\big).
 \label{solI}
\end{equation}
whose Cotton-York tensor in Jordan form
is given by Eq. \eqref{Jsol} which makes it algebraically general. The theory fixes the characteristic length scale $\ell$ by
\begin{equation}
 \ell^2=\frac{1}{2\Lambda}.
\end{equation}
Moreover, the theory's coupling constants are constrained by
\begin{equation}
 m^2=5\Lambda>0,
\end{equation}
which should be compared to Eqs. \eqref{sol1} and \eqref{sol2}.

Similarly, another algebraically general Sol geometry is given by
\begin{equation}
 g_{I'}=\ell^2\big(-e^{-2z}\d x\otimes\d x +e^{2z}\d y\otimes\d y+\d z\otimes\d z\big).
 \label{solI'}
\end{equation}
This time, the Jordan form of the Cotton-York tensor has the form
\begin{equation}
Y = 2\ell^3
\left(\begin{array}{ccc}
 0 & 1 & 0 \\
 -1 & 0 & 0 \\
 0 & 0 & 0
\end{array}\right).
\label{Jsol'}
\end{equation}
Note that the difference between Eqs. \eqref{solI} and \eqref{solI'} is that the corresponding eigenvalues of Eqs. \eqref{Jsol} and \eqref{Jsol'} are real and purely imaginary, respectively.
Much like Minkowski and AdS spacetimes, this Lorentzian geometry obeys NMG dynamics under the same conditions as its Euclidean counterpart, i.e., Eqs. \eqref{sol1} and \eqref{sol2}.

The third Sol geometry under consideration is
\begin{equation}
 g_{N}=\ell^2\big(-e^{2z}\d y\otimes\d y-2\d x\otimes_s\d y +\d z\otimes\d z\big),
 \label{solN}
\end{equation}
for which 
\begin{equation}
Y = 
\left(\begin{array}{ccc}
 0 & 1 & 0 \\
 0 & 0 & 0 \\
 0 & 0 & 0
\end{array}\right),
\end{equation}
indicating that it is algebraically type $N$.
At a first glance, the metric in Eq. \eqref{solN} can be seen to admit a null vector field
\begin{equation}
 V=\frac{\partial}{\partial x}.
\end{equation}
This field is covariantly constant, hence, the metric describes a pp-wave spacetime.
The NMG tensor is
\begin{equation}
 K(g_N)=\frac{8}{\ell^2}\ric(g_N)=\frac{16}{\ell^2}e^{2z}\d y\otimes\d y,
\end{equation}
so for this pp wave to be a vacuum solution it requires
\begin{equation}
 \Lambda=0,\quad\text{and}\quad m^2>0,
\end{equation}
as
\begin{equation}
 \ell^2=\frac{4}{m^2}.
\end{equation}
Since this solution is type $N$, it must belong the general family found in~\cite{Ahmedov:2010uk}. However, as a specific geometry, it represents a new solution of NMG. Moreover, it is the only pp-wave vacuum solution which has Sol isometry.

To close this section, let us remark that Eqs. \eqref{Lorentz-Sol}, \eqref{solI'} and \eqref{solN}, which are new, together with Eq. \eqref{solI} which was discussed in~\cite{Ahmedov:2010em,Ahmedov:2011yd}, all have different 
Petrov type.

\emph{Closing Remarks.}---We conclude by mentioning that
early work on Thurston geometries as Euclidean gravitational configurations came from string theory~\cite{Gegenberg:2002xj}. A stringy 3D gravity theory was constructed as a way to shed light on to the Thurston conjecture~\cite{Gegenberg:2003yz}.
A key feature of that work is that the gravity sector is given by general relativity. Since most of the Thurston geometries are not Einstein manifolds, then in previous analyses a unifying dynamics required matter content. In this Letter, unifying dynamics comes from modifying Einstein gravity. In other words, we do not consider matter fields. This establishes that all eight Thurston geometries are NMG vacua. To the best of our knowledge, NMG is the only 3D gravity theory where this happens.

We thank Eloy Ay\'on-Beato for suggesting to us we explore Thurston geometries in new massive gravity. We are also grateful to Mokhtar Hassa\"ine for helpful discussions. D. F.-A. acknowledges financial support from CONACYT through a postdoctoral research grant. This work has been partially funded by CONACYT Grant No. A1-S-11548.

\bibliography{Thurston}

\end{document}